\documentclass{ws-procs9x6}
\usepackage{pennames}

\def\Journal#1#2#3#4{{#1} {\bf #2}, #3 (#4)}

\def\PLB{{\em Phys. Lett.}  B}

\def\ifb{\ensuremath{\mathrm {fb}^{-1}}}
\def\pb{\ensuremath{\mathrm {pb}}}

\def\Z{\mbox{$\mathrm{Z}$}}

\def\MW{\mbox{$m_{\mathrm{W}}$}}

\def\GW{\mbox{$\Gamma_{\mathrm{W}}$}}

\def\GEV{\mbox{$\mathrm{GeV}$}}

\def\GEVcc{\mbox{$\mathrm{GeV}/{{\it c}^2}$}}

\def\epem{\mbox{$\mathrm{e}^+\mathrm{e}^-$}}
\def\ee{\mbox{$\mathrm{e}\mathrm{e}$}}
\def\rs{\mbox{$\sqrt{s}$}} 
\def\qq{\mbox{$\mathrm{qq}$}}
\def\qqbar{\mbox{$\mathrm{q\overline{q}}$}}

\def\vvbar{\mbox{$\nu\overline{\nu}$}}
\def\ev{\mbox{$\mathrm{e}\nu$}}
\def\mv{\mbox{$\mu\nu$}}
\def\tv{\mbox{$\tau\nu$}}
\def\lv{\mbox{$\ell\nu$}}

\def\ra{\rightarrow}

\begin{document}

\title{Electroweak Physics at LEP2}

\author{P. AZZURRI}

\address{Scuola Normale Superiore, \\
Piazza dei Cavalieri 7, \\ 
56126 Pisa, Italy\\ 
E-mail: p.azzurri@sns.it}

\maketitle

\abstracts{
The measurements resulting from the 
analysis of the LEP2 data have brought more strong 
evidence in support of the standard electroweak model.
In particular the LEP2 data has revealed 
(i) the first determination of the SU(2) gauge bosons
self-couplings, (ii) the first direct measurements of
the W decay-couplings, and (iii) the current best direct
measurement of the W mass. 
}

\section{Introduction}
During the LEP2 program a total of about 3~\ifb
of $\epem$ data at centre-of-mass energies 
$\rs$=161-209~$\GEV$, have been collected.
A selection of Standard Model (SM) electroweak measurements 
established with this data is given in the following.

\section{Single Photons and Photon Pairs}

\begin{figure}[hbt]
\centerline{\epsfxsize=2.0in\epsfbox{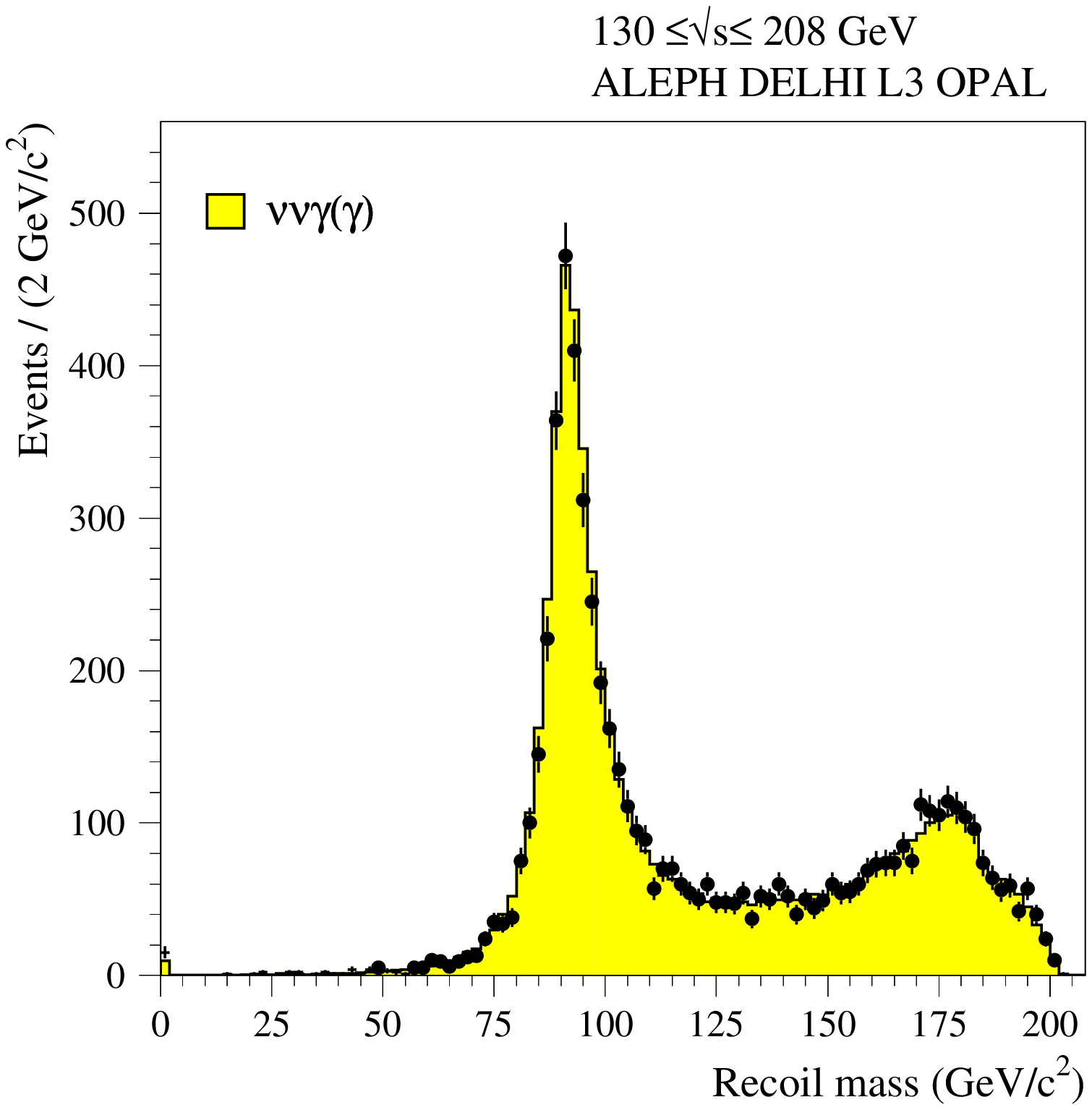}
  \epsfxsize=3in\epsfbox{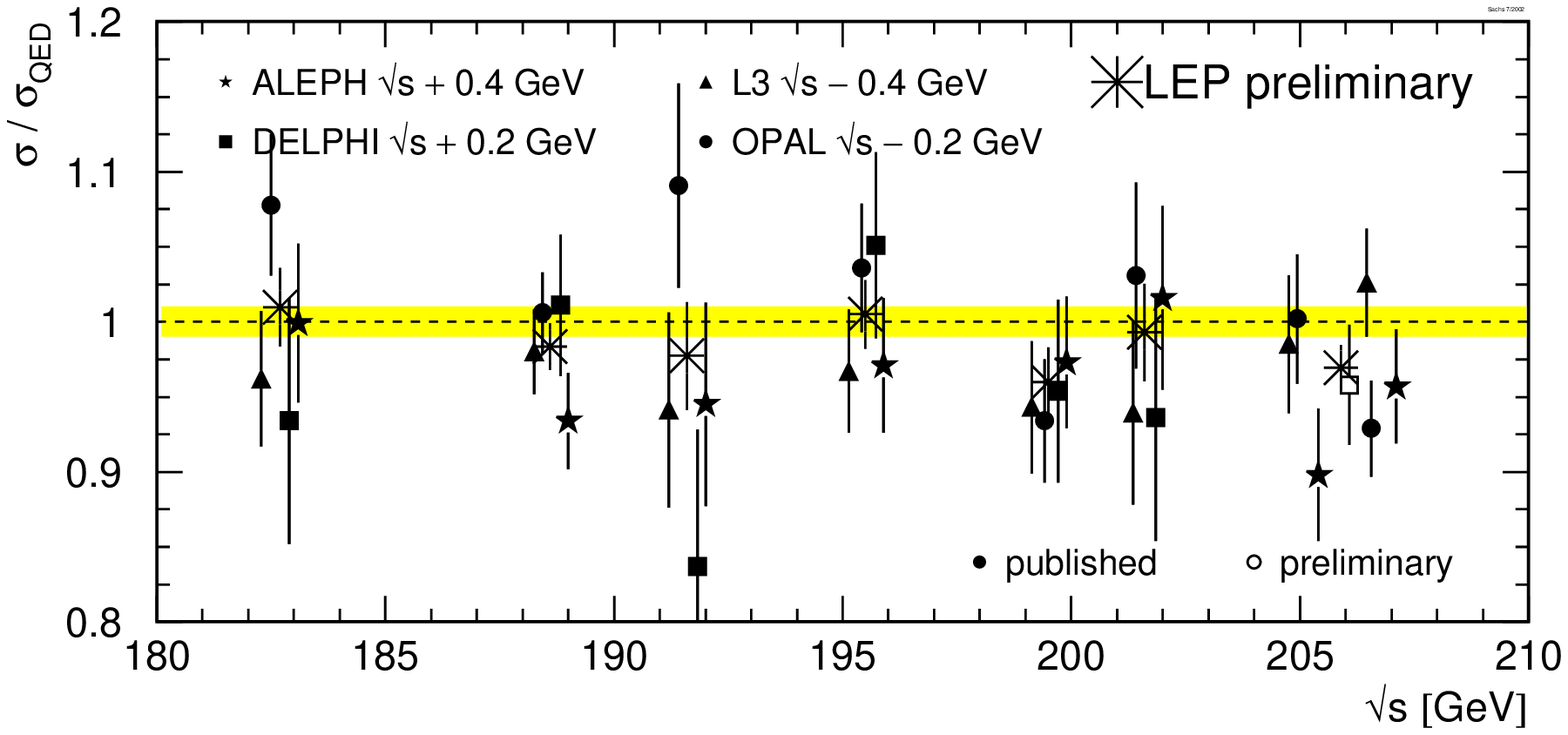}}   
\caption{
Distribution of the recoil mass in single photon events
from all LEP2 data (left).
Ratios of measured cross-section over QED predictions
for photon-pair production at different LEP2 energies. The shaded
area represents the theoretical uncertainty (right).
\label{fig:phot}}
\end{figure}

Single photon final states arise at LEP2 from the 
$\epem\ra\Z\gamma\ra\vvbar\gamma$ process.
The missing mass distribution in single photon events
selected at LEP2 energies is
shown in Fig.~\ref{fig:phot}, 
and clearly shows the Z mass peak
decaying into neutrino pairs. 
The analysis of these events provides a 
direct measurement of the Z decay rate to neutrinos, and from this 
the number of light neutrino families is derived to be 
$ N_\nu=2.84\pm 0.08$, in good agreement with the more precise 
indirect determination from the LEP1 Z width measurements,
 $ N_\nu=2.984\pm 0.008$\cite{pdg04}.

Photon pair productions provide a test of the purely QED 
process $\epem\ra\gamma\gamma$. As can be seen in Fig.~\ref{fig:phot},
the measured LEP2 cross-sections agree nicely with the QED 
predictions at the percent level.

Both results on single photon and photon-pair productions at LEP2
can be used to extract limits on the scale of many 
new physics models beyond the Standard Model, 
up to the TeV level\cite{ew}.

\section{Fermion Pairs}

Fermion pairs with $\qqbar$, $\Pgmp\Pgmm$ and $\Pgt^+\Pgt^-$ 
final states are produced in $\epem$ collisions with a $\gamma$/Z 
$s$-channel exchange. The total cross-sections and forward-backward
asymmetries measured at LEP2 energies are
in agreement with the electroweak predictions for $\gamma$/Z interference
at the percent level. 
Bhabha $\epem\ra\epem$ final states get additional 
large contributions from the 
$t$-channel $\gamma$ exchange, for forward angle scattering. 
Also in the Bhabha channel the LEP2 measured total and differential 
cross-sections are in nice agreement with electroweak predictions
at the 1-10\% precision level, according to the scattering angle\cite{ew}.

Just like for photons, also fermion pair final states 
have been analyzed to extract limits
on the scale of different
physics models beyond the Standard Model, 
up to the TeV level\cite{ew}.

\section{Single W and Z}
Single electroweak boson productions $\epem\ra\PW\ev$ and
$\epem\ra\Z\ee$ represent four-fermion final states. 
The luminosity weighted cross-section averages for
these processes,
at the LEP2 average centre-of-mass energy of $\rs\simeq$198~GeV,
are $\sigma(\epem\ra\PW\ev)= 0.77 \pm 0.05 ~\pb $ 
for single W production, and 
$ \sigma(\epem\ra\Z\ee\ra\qq\ee)= 0.55 \pm 0.03 ~\pb $
for single Z production. 
Both measurements are in agreement with the SM 
expectations at 7\% and 5\% precision level\cite{ew}.

\section{W and Z Pairs}
W-pair production is one of the most interesting processes
of the electroweak model, where the 
non-abelian structure of the SU(2)
group leads to the presence of gauge boson self couplings
that play a crucial r\^ole establishing the gauge cancellations
that guarantee the W-pair process unitarity and 
the renormalizability 
of the theory.

Results for the W-pair cross-sections as a function of
the LEP2 energy\cite{ew} are shown in Fig.~\ref{fig:wz}, and are in
agreement with the SM expectations at the 1\% level. 
The measured W-pair production rates represent the first clear proof of the 
presence of both the WW$\gamma$ and WWZ couplings dictated by the 
SU(2)$\otimes$U(1) gauge structure.

Results for LEP2 Z-pair cross-sections are also shown 
in Fig.~\ref{fig:wz} where the agreement with the SM expectations
is at the level of 5\%.
In this case the Z-pair data rules out the presence of 
purely neutral gauge self couplings, such as ZZ$\gamma$ and ZZZ vertices,
that are not predicted by the 
electroweak theory.

\begin{figure}[hbt]
\centerline{\epsfxsize=2.1in\epsfbox{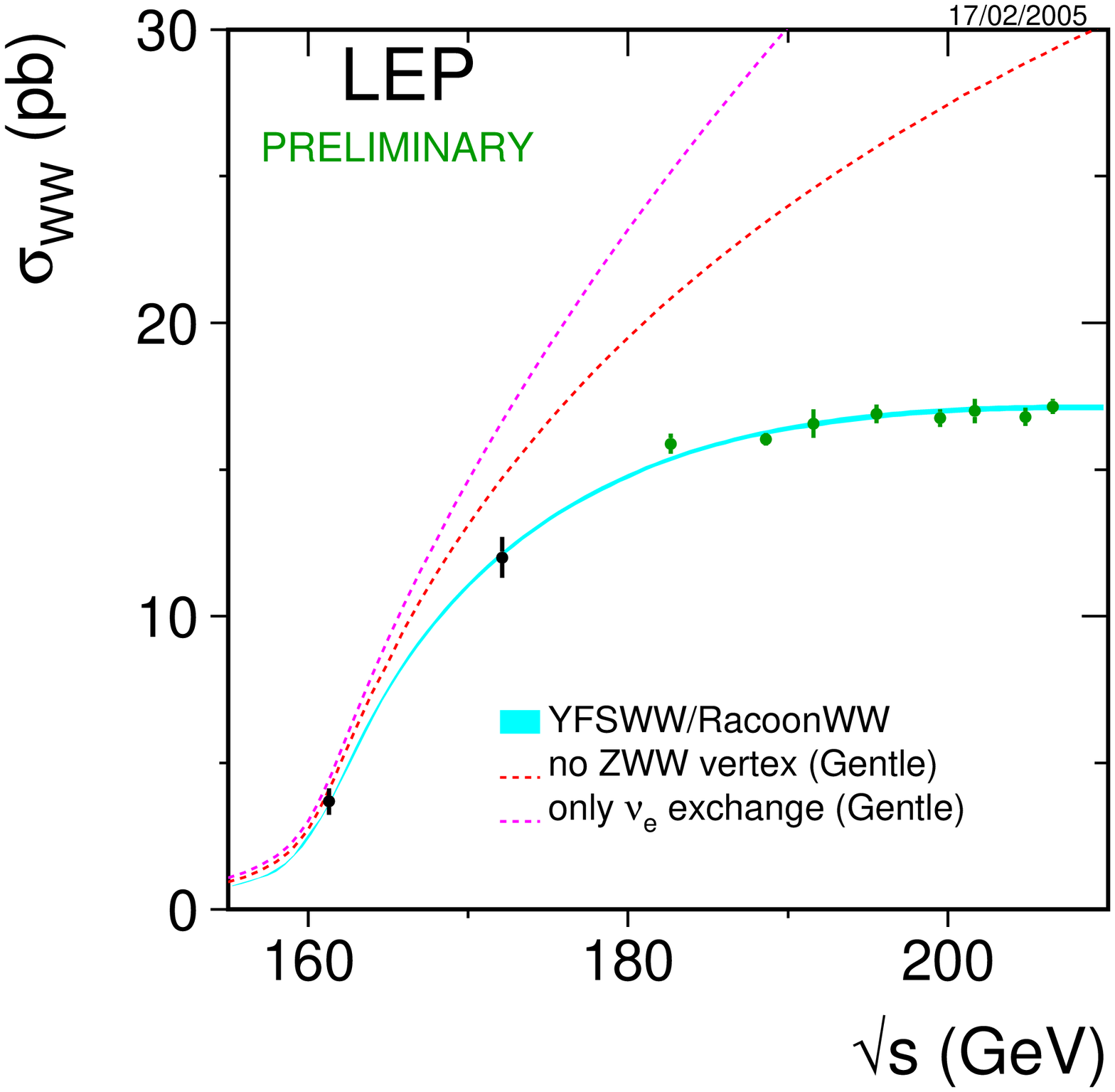}
\epsfxsize=2.1in\epsfbox{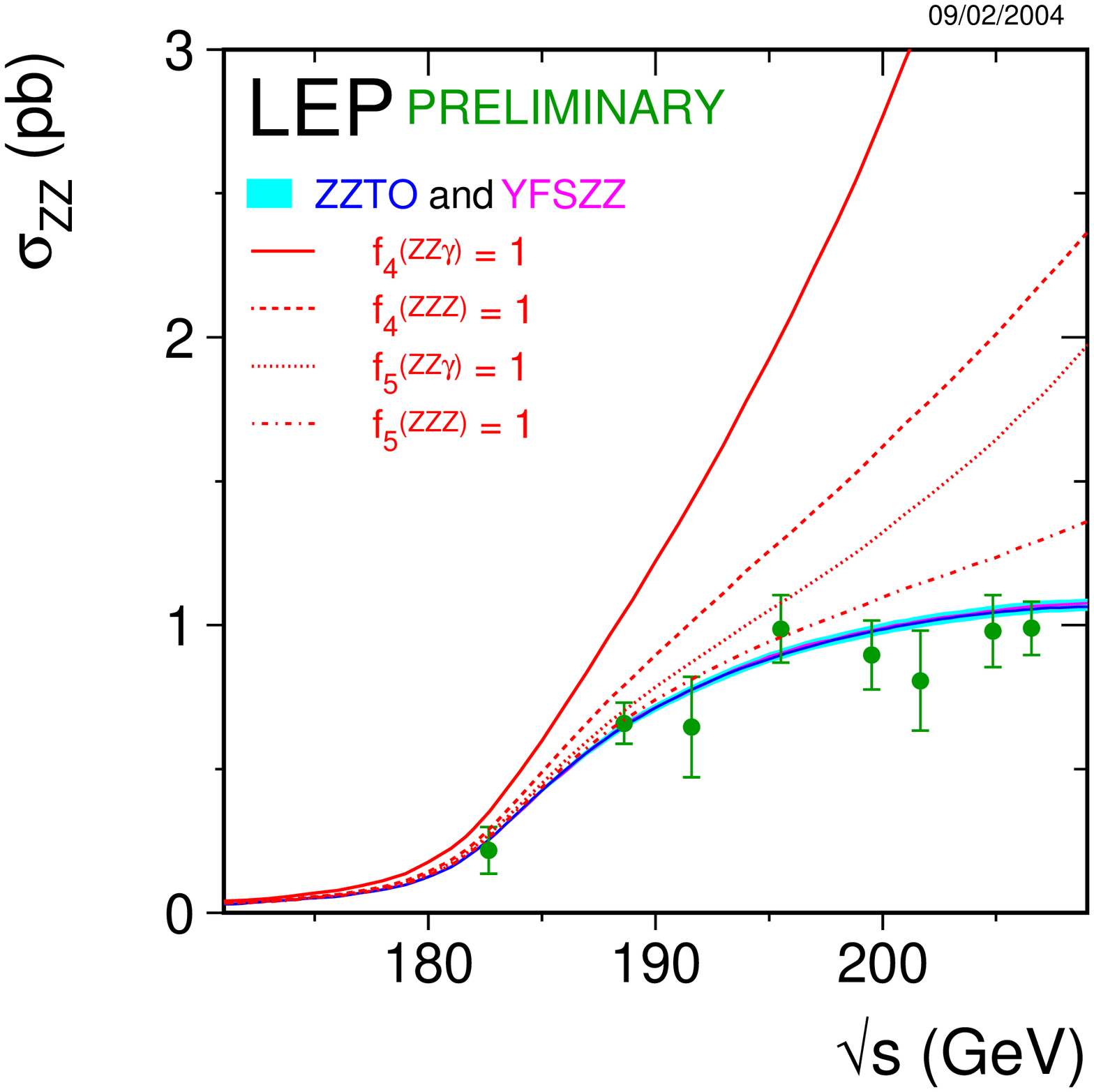}}   
\caption{
Combined LEP results for production cross-sections 
for W-pairs (left) and Z-pairs (right), as a function of the 
centre-of-mass energy. The lower shaded curves represent the 
Standard Model predictions and uncertainties.
The upper curves on the W-pair production plot show
the predictions in the absence of the $\gamma$WW and ZWW 
couplings.  The upper curves on the Z-pair production plot show
the predictions in the presence of different possible 
ZZ$\gamma$ and ZZZ couplings.
\label{fig:wz}}
\end{figure}

\section{Gauge bosons self-couplings}
The structure and magnitude of the $\gamma$WW and ZWW couplings
are extracted from the W-pair event rates and angular 
distributions\cite{ew}. 
A fit with the ALEPH data\cite{atgc} to the 28 parameters of the most 
general Lorentz-invariant vertex structures leads to results
in agreement with the SU(2)$\otimes$U(1) predictions with 
precisions of 3-20\%, according to the measured parameter.

A more constrained fit of all LEP2 data,
in search of anomalous contributions in gauge couplings,
 to the three couplings that conserve 
separately C and P, U(1)$_{\rm em}$, and global SU(2)$\otimes$U(1),
yields\cite{ew}
$\kappa_{\gamma}= 0.984 \pm 0.045$,
$\lambda_{\gamma}= -0.016 \pm 0.022$, and
$g_1^{\rm Z}= 0.991 \pm 0.021$,  
revealing again no deviation from the SM expectations.

\section{W decay couplings}
The LEP2 W-pair sample has allowed the first direct
measurements of all hadronic and leptonic W decay
branching ratios to be 
${\rm B}(\PW\ra\ev)= 10.69 \pm 0.17\%$, 
${\rm B}(\PW\ra\mv)= 10.57 \pm 0.16\%$,
${\rm B}(\PW\ra\tv)= 11.39 \pm 0.23\%$, and
${\rm B}(\PW\ra\qq )= 67.51 \pm 0.29 \%$.
These results insure the lepton-quark universality of charged currents
at the 0.6\% level ($g_{\rm q}/g_\ell= 1.000\pm 0.006$),
and of the lepton family universality of charged currents at the 
1\% level.
However, the tau coupling to the W appears to be 2.6
standard deviations larger than the combined electron and muon couplings
as $2g_\tau/(g_{\rm e}+g_\mu)= 1.036\pm 0.014$.

The W hadronic decay fraction can also be interpreted as a test
of the unitarity of the CKM quark mixing matrix in the 
first two families, as 
$\sum |V_{ij}| (i=u,c ;\; j=d,s,b)= 2.000\pm 0.026$, and from 
this extract the W$cs$ coupling amplitude 
$|V_{cs}|=0.976\pm 0.014$,
without CKM unitarity assumptions. 

\section{W boson mass and width}
The first LEP2 W mass determination has been extracted  
from the  W-pair production threshold cross-section\cite{ew}
yielding $ \MW=80.40\pm 0.20 ~\GEVcc $.
For the direct measurement,
the W invariant mass is reconstructed event-by-event
in all $\qq\qq$ and $\qq\lv$ decays of W-pairs,
from the kinematics of the visible decay particles, and
the resolution of the W mass peak is improved 
by applying a kinematic fit imposing 
energy-momentum conservation constraints from the LEP2 energy.
The W mass and width values
can be extracted from the W mass data distributions
using different fit methods, and yielding 
$\MW=80.412\pm 0.042 ~\GEVcc $, and 
$\GW=2.150\pm 0.091 ~\GEVcc $, where the weight 
of the fully hadronic ($\qq\qq$) channel
is only 10\% because of large uncertainties coming from 
possible final state interactions 
effects between 
the two W hadronic decay products.
The inclusion of the current W mass determinations in 
the electroweak
fit yields a constraint on the SM higgs mass 
$114<m_h<260\GEVcc$ at 95\% confidence level\cite{ew}.


\begin{figure}[hbt]
\centerline{\epsfxsize=2.1in\epsfbox{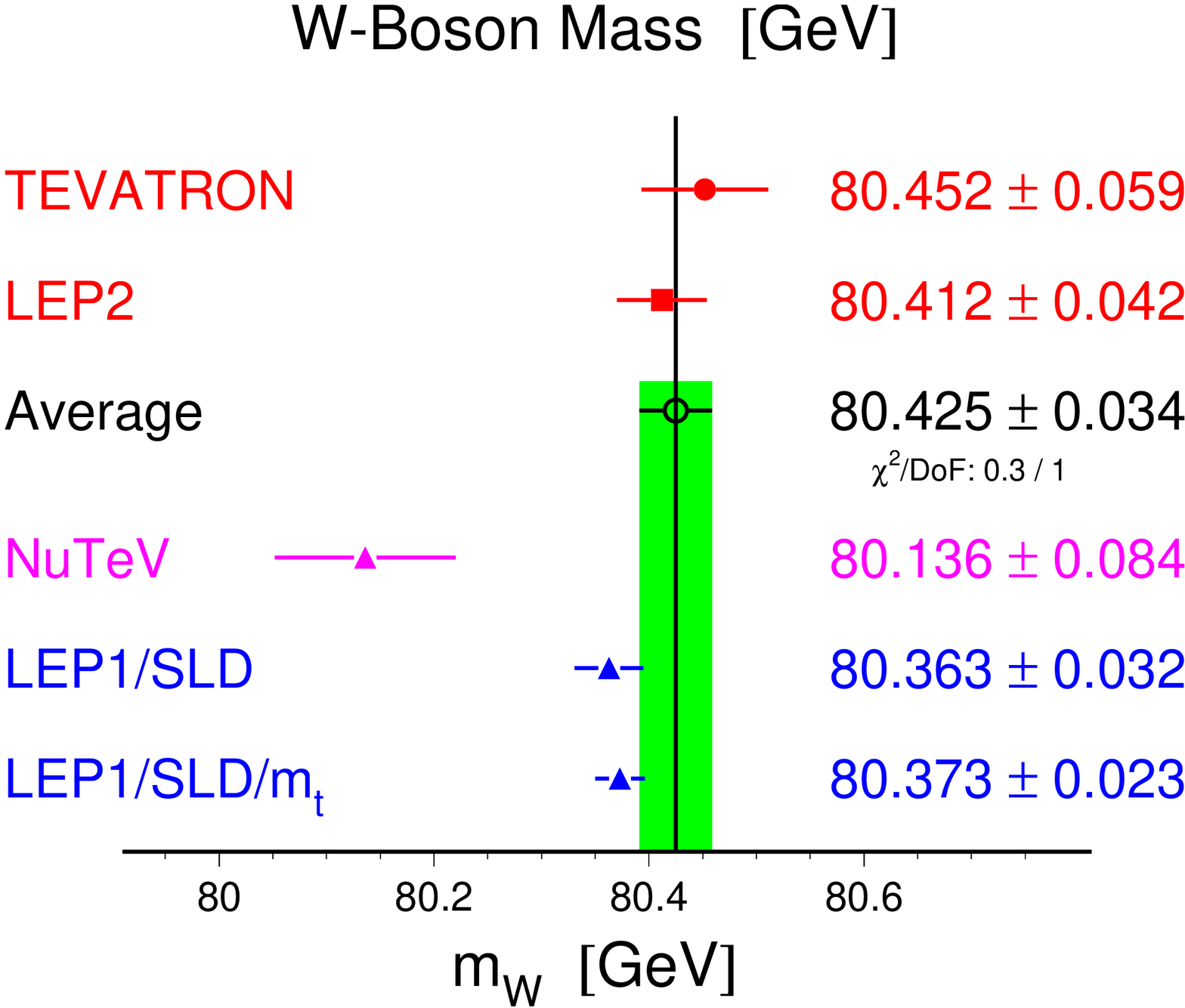}
\epsfxsize=2.1in\epsfbox{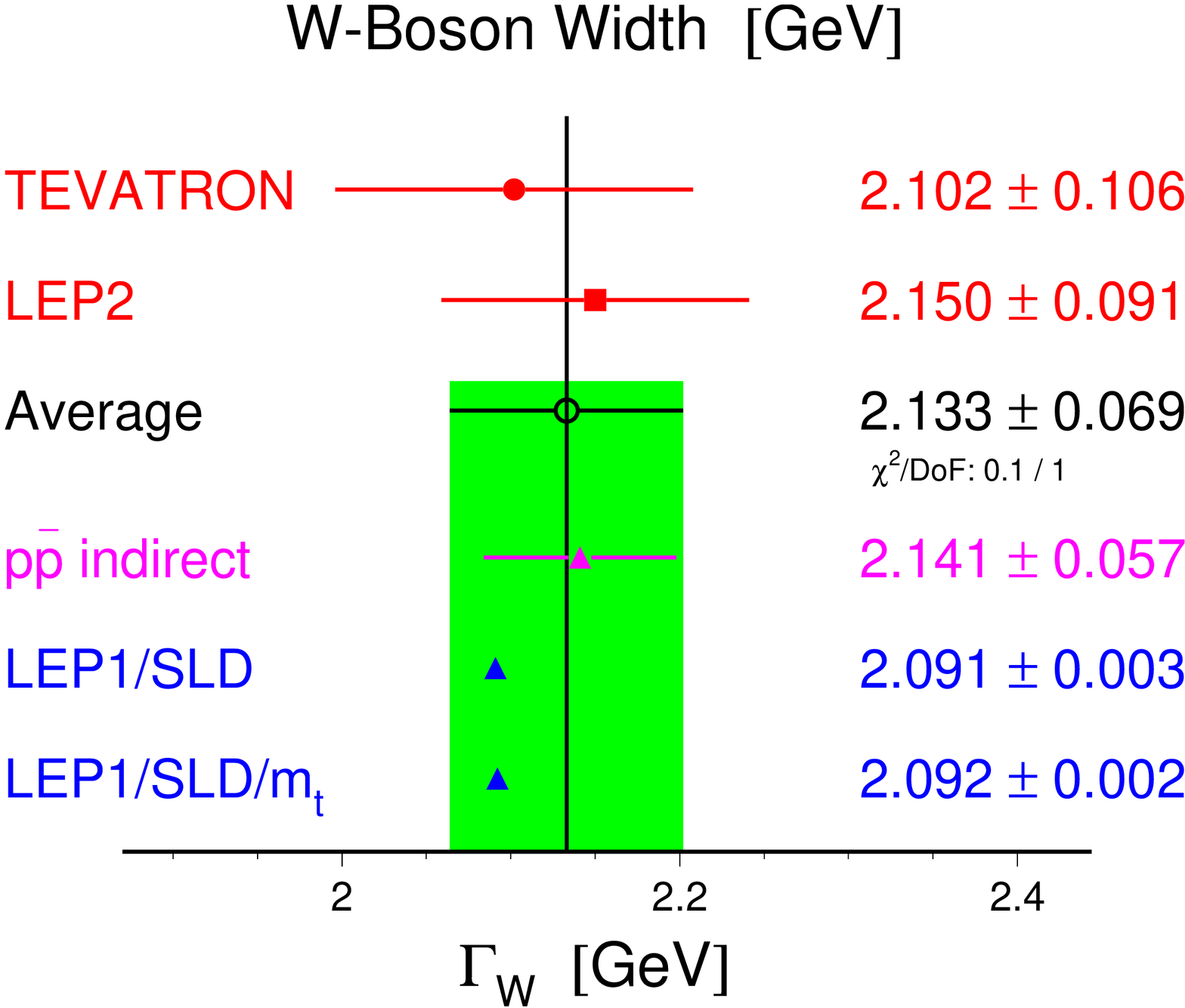}}   
\caption{
Summary of W mass and width measurements. 
Direct measurements from LEP2 and the TEVATRON are shown on the top,
indirect constraints from other electroweak determinations on
the bottom. 
 \label{fig:mw}}
\end{figure}

\section*{Acknowledgments}
I would like to thank Roberto Chierici for providing the most recent 
combined results for four-fermion productions. 
I also have to thank Stefania who fortunately could not come to the 
conference, and Aafke Kraan for nicely reviewing this paper.

\end{document}